\documentclass[lettersize,journal]{IEEEtran}
\usepackage{amsmath,amsfonts}
\usepackage{algorithmic}
\usepackage{algorithm}
\usepackage{array}
\usepackage[caption=false,font=normalsize,labelfont=sf,textfont=sf]{subfig}
\usepackage{textcomp}
\usepackage{stfloats}
\usepackage{url}
\usepackage{verbatim}
\usepackage{graphicx}
\usepackage{cite}
\usepackage{makecell}

\begin{document}

\title{Holistic Grid-Forming Control to Enhance the Frequency Support from HVDC-Connected Offshore Wind Power Plants}

\author{Zhenghua~XU,~\IEEEmembership{Student~Member,~IEEE,}
Dominic~Groß,~\IEEEmembership{Senior~Member,~IEEE,}
George~Alin~Raducu,
Behnam~Nouri,~\IEEEmembership{Member,~IEEE,}
Oscar~Saborío-Romano,~\IEEEmembership{Senior~Member,~IEEE,}
and Nicolaos~A.~Cutululis,~\IEEEmembership{Senior~Member,~IEEE}
\thanks{This work was supported by the ADOreD project of the European
Union’s Horizon Europe Research and Innovation Program under the Marie Skłodowska-Curie Grant 101073554. \textit{(Corresponding
author: Zhenghua Xu.)}}
\thanks{Zhenghua~XU, Oscar~Saborío-Romano, and Nicolaos~A.~Cutululis are with the Department of Wind and Energy Systems, Technical University of Denmark, Roskilde 4000, Denmark (email: zhexu@dtu.dk; osro@dtu.dk; niac@dtu.dk).

Dominic~Groß is with the Department of Electrical and Computer Engineering, University of Wisconsin-Madison, Madison, WI 53706, USA
(e-mail: dominic.gross@wisc.edu).

Zhenghua~XU and George~Alin~Raducu are with Vattenfall Vindkraft A/S, Kolding 6000, Denmark (email: zhenghua.xu@vattenfall.com; alingeorge.raducu@vattenfall.com).

Behnam~Nouri is with Vattenfall Winkraft Europe GmbH, Hamburg 20457, Germany (email: behnam.nouri@vattenfall.de).
}}

\markboth{Journal of \LaTeX\ Class Files,~Vol.~14, No.~8, August~2021}%
{Shell \MakeLowercase{\textit{et al.}}: A Sample Article Using IEEEtran.cls for IEEE Journals}


\maketitle

\begin{abstract}
To address the frequency stability challenges posed by the rising penetration of power electronics in power systems, HVDC-connected offshore wind power plants (OWPPs) are increasingly expected to provide inertial response and frequency containment reserve (FCR). In this paper, an improved holistic grid-forming (GFM) control is proposed, aiming to enhance the frequency support by coordinating the GFM controls implemented at all AC and DC terminals of an HVDC-OWPP system, without requiring communication. Firstly, the model of a typical HVDC-OWPP system is developed for control design. Accordingly, the proposed controllers are formulated, followed by an analytical tuning method, where the upper bound of the bandwidth at each AC or DC terminal is identified. Finally, simulations are conducted to verify the functionality and compare the performance with that of representative control configurations. The results show that the proposed holistic GFM control achieves faster response and thus more effective frequency support, while the utilization of the inherent energy storage of each converter is minimized, thereby supporting a new design philosophy for converter control in converter-dominated systems.


\end{abstract}


\section{Introduction}
\IEEEPARstart{T}{he} ongoing energy transition is driving the growing integration of converter-based renewable power generation into power systems. As many fossil-fuel plants are being phased out, the associated reduction in the frequency support traditionally provided by these units, such as inertia and regulation reserves, is placing increasing stress on system frequency stability~\cite{NJOKA20244983}. As an emerging large-scale power supply, HVDC-connected offshore wind power plants (OWPPs) are expected to provide ancillary services to support the onshore grid frequency. Moreover, participation in ancillary service markets also offers OWPPs an additional revenue stream to mitigate the impact of diminishing policy incentives~\cite{noauthor_new_2018}. Among the ancillary services for system frequency stability, inertial response plays a key role in limiting the rate of change of frequency (RoCoF) after an event until frequency containment reserve (FCR) can arrest frequency deviations~\cite{10158921}. Therefore, this paper investigates controls for HVDC-connected OWPP (HVDC-OWPP) that provide an inertial response and FCR to the onshore grid. In particular, we focus on modular
multilevel converter (MMC) based HVDC system and OWPPs consisting of type-4 wind turbine generators (WTGs) with 2-level voltage source converters (VSCs).

Existing control approaches can be categorized into communication-based control (CBC) and communication-free control (CFC). Under CBC~\cite{5223560}, the onshore frequency is measured at the onshore station and then transmitted through a communication network to the offshore side so that OWPPs can respond accordingly. Under CFC~\cite{6107607}, the onshore frequency deviation is incorporated into the onshore DC voltage reference to be reflected in the DC voltage variation. In parallel, the offshore frequency is modulated based on the offshore DC voltage deviation for OWPPs to respond. 

Performance comparisons between CBC and CFC have been conducted in~\cite{9428500,9124711,9507468}, demonstrating that CFC generally achieves less oscillations and better frequency deviation suppression than CBC for no communication-delay. Therefore, this paper focuses on CFC. However, in these comparisons, converter control configurations are limited, leaving room for exploring alternative configurations of converter controls which may
deliver improved performance and robustness.

Existing converter controls can generally be categorized into grid-following (GFL) and grid-forming (GFM) control. At the AC terminal, AC-GFL converters typically rely on a phase-locked loop (PLL) to synchronize with the AC grid. In contrast, AC-GFM converters may adopt active power~\cite{5308285}, DC voltage~\cite{8011460}, or the internal energy of the MMC~\cite{9887878} for synchronization. At the DC terminal, DC-GFM converters regulate the terminal DC voltage, whereas DC-GFL converters do not. Therefore, converter control can be further categorized into four types: AC-GFL \& DC-GFM, AC-GFM \& DC-GFL, dual-port GFL, and dual-port GFM control~\cite{9729620}.



A comparison of the frequency support performance of AC-GFL and AC-GFM controls has been conducted in~\cite{zuo_performance_2021,8586162}, where AC-GFM control presents significant advantages, achieving lower frequency excursion and RoCoF. At the same time, the comparison between DC-GFL and DC-GFM controls, as well as the interaction between the converter AC terminal, internal energy storage, and DC terminal remains largely unexplored.

The converter control configuration for an HVDC-OWPP system has undergone rapid evolution. Initially, both CBC~\cite{7094249} and CFC~\cite{6345609} employed AC-GFL \& DC-GFM control for the onshore station, AC-GFM \& DC-GFL control (specifically, $V-f$ GFM control) for the offshore station, and AC-GFL \& DC-GFM control for the grid-interface of WTGs. In this classic configuration, the presence of GFL controlled converters breaks the inherent physical response chain, causing insufficient coordination between AC and DC dynamics. Thus, GFM control has gradually been introduced across  HVDC-OWPPs to improve their frequency support capabilities. 

GFM control of type-4 WTGs in OWPPs is reviewed in~\cite{9589838,huang2023challenges}. Notably, using virtual synchronous generator (VSG) control (AC-GFM \& DC-GFL) on the grid-side converter (GSC) and DC-GFM control on the machine-side converter (MSC) results in  DC voltage source behavior of the MSC and enables inertial response capability of the GSC. However, this configuration is not compatible with maximum power point tracking (MPPT) and may result in excessive mechanical stress on the WT. To tackle this issue, dual-port GFM control can be used for the GSC and further improved by being coordinated with machine-side controls~\cite{10091242,10734179}. 

For the HVDC system, GFM control is typically used under the CFC paradigm. For instance, \cite{8950158,9887878,10532701} proposed various dual-port GFM controls for the onshore MMC, while the offshore MMC always operates under $V-f$ GFM control which can only transfer power response passively and lacks coordination capability. 
Then, the internal energy of the offshore MMC is incorporated into the frequency response through offshore DC-GFL control~\cite{10044241}. Moreover, dual-port GFM control on the offshore MMC is combined with AC-GFM \& DC-GFL control on the onshore MMC~\cite{10926727,10916051}, where DC-GFL control remains.
Finally, dual-port GFM control for both onshore and offshore MMC has been shown to achieve full frequency response coordination for lossless networks~\cite{9729620,10478164}.

Studies that jointly consider the HVDC system and OWPPs have shown significant advantages of using GFM controls for all AC terminals of an HVDC-OWPP system (i.e., full AC-GFM control). For instance, in comparison to the classic configuration, full AC-GFM results in better DC-side damping, stable operation under weak onshore and offshore AC grids, and faster frequency support~\cite{9962793}. In addition, a comparison between a full AC-GFM control, a partial AC-GFM control, and the classic configuration demonstrated that the full AC-GFM strategy exhibits more flexibility~\cite{10926727}.

In summary, although GFM control has been leveraged to improve the frequency support capabilities of HVDC-OWPP systems, studies that address full deployment of GFM control at all AC and DC terminals of an HVDC-OWPP system remain limited. Notably, deploying dual-port GFM control on all converters only results in full synchronization and coordination of inertia response and FCR for lossless networks~\cite{10478164}. In contrast, holistic GFM control achieves full synchronization and frequency response coordination~\cite{11180312}, but its dynamic performance and control tuning remains unexplored.

In this paper, an improved holistic GFM control is proposed to enhance the inertial response and FCR provision from an HVDC-OWPP system. Moreover, a control tuning method is developed, based on the loop-shaping analysis among the AC and DC terminals. In particular, maximizing the converter control bandwidth results in faster frequency response and improved provision of inertia response and FCR from the OWPP, while decreasing the reliance on the internal energy storage of converters. The results are verified using time-domain simulations. 


\section{System Modeling}
\begin{figure*}[!t]
\centering
\subfloat[]
{\includegraphics[width=\textwidth]{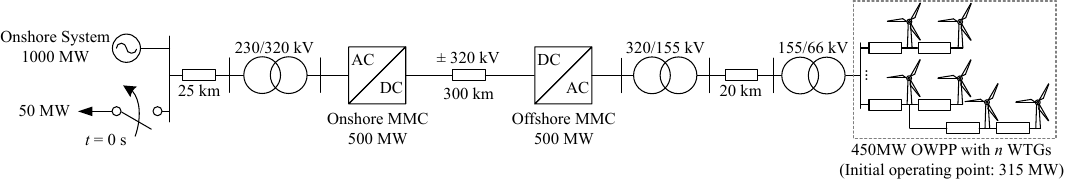}%
\label{fig_SysScheme}}
\hfil
\subfloat[]{\includegraphics[width=\textwidth]{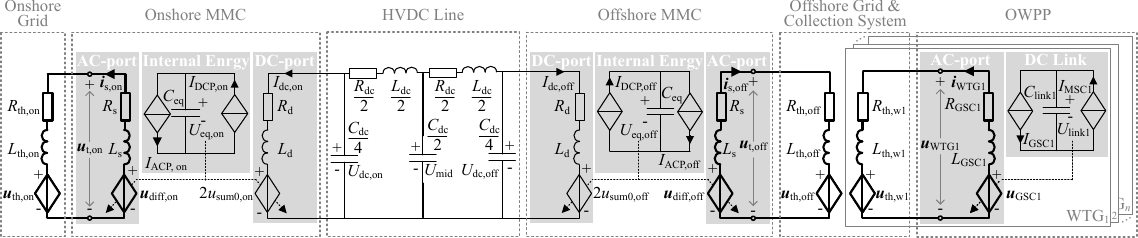}%
\label{fig_EquivCircuit}}
\hfil
\subfloat[]{\includegraphics[width=\textwidth]{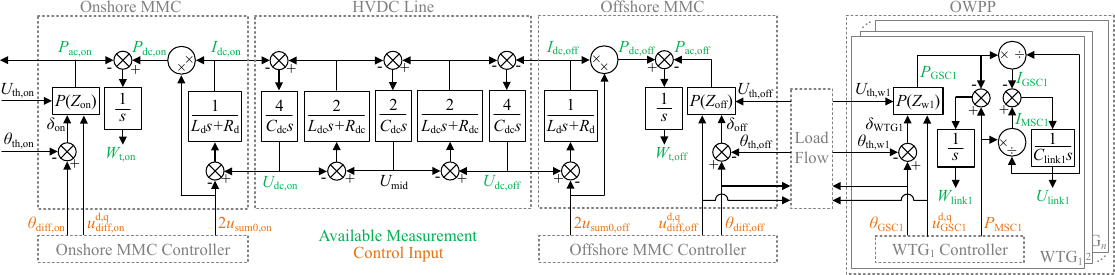}%
\label{fig_ControlBox}}
\caption{A typical point-to-point HVDC-OWPP system. (a) Single-line schematic diagram. (b) Equivalent circuit (three-phase systems are in bold). (c) Block diagram of the plant}
\label{fig_SystemModel}
\end{figure*}

A point-to-point HVDC-OWPP system as shown in Fig.~\ref{fig_SysScheme} is studied in this paper. To inform the control design, this section develops a model of the HVDC–OWPP using a component-by-component approach, specifically addressing the MMCs, OWPP, and HVDC line, while the onshore and offshore AC grids are incorporated through the AC terminal dynamics of the MMCs and OWPP. For each component, the electrical properties are captured in the equivalent circuit shown in Fig.~\ref{fig_EquivCircuit}. Subsequently, a corresponding block diagram (see Fig.~\ref{fig_ControlBox}) is derived from the equivalent circuit.

\subsection{Onshore \& Offshore MMCs}
\begin{figure}[!t]
\centering
\includegraphics[width=0.7\columnwidth]{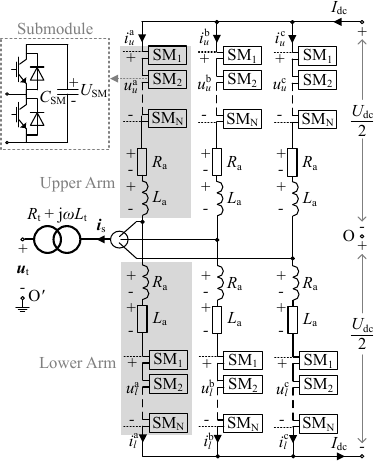}
\caption{Generic monopole MMC with half-bridge submodules}
\label{fig_MMC}
\end{figure}

A generic monopole topology with $N$ half-bridge submodules (SMs) in each arm is considered for both onshore and offshore MMCs, as shown in Fig.~\ref{fig_MMC}, where $\textit{\textbf{u}}_{\text{t}}$ (V) is the phase-to-ground terminal voltage; $R_{\text{t}}$ ($\Omega$) and $L_{\text{t}}$ (H) are the winding resistance and leakage inductance of the transformer; $\omega$ (rad/s) is the actual angular frequency of the AC system, and $\omega = 2\pi f$, where $f$ (Hz) is the frequency of the AC system; $\textit{\textbf{i}}_{\text{s}}$ (A) is the output current to the connected AC system; $R_{\text{a}}$ ($\Omega$) and $L_{\text{a}}$ (H) are the arm resistance and arm inductance; $u^{j}_{u,l}$ (V) is the total inserted submodule voltage of an upper or lower arm of phase \textit{j}, and $j \in {\{\text{a, b, c}\}}$; $i^{j}_{u,l}$ is the current (A) flowing though a upper or lower arm of phase \textit{j}, and $j \in {\{\text{a, b, c}\}}$; $U_{\text{dc}}$ (V) and $I_{\text{dc}}$ (A) are the DC voltage and current.

\subsubsection{MMC AC terminal}
Focusing on the fundamental frequency component and assuming the symmetrical operation of the AC system, the dynamics of $\textit{\textbf{i}}_{\text{s}}$ in Fig.~\ref{fig_MMC} can be modeled by adding the voltage balance equations of upper and lower arms. Using
$L_{\text{s}} = L_{\text{t}} + L_{\text{a}} / 2$; $R_{\text{s}} = R_{\text{t}} + R_{\text{a}} / 2$;  
$u^{j}_{\text{diff}}$ is the differential voltage of phase $j$, and $u^{j}_{\text{diff}} = ( u^{j}_{\text{l}} - u^{j}_{\text{u}} ) / 2$, we obtain
\begin{equation}
\label{MMC_OutputCurrent2}
L_{\text{s}} \frac {\text{d}i^{j}_{\text{s}}} {\text{d}t} 
+ R_{\text{s}} i^{j}_{\text{s}} 
= 
u^{j}_{\text{diff}}
- u^{j}_{\text{t}}, 
\quad
\forall j \in {\{\text{a, b, c}\}}.
\end{equation} 
Therefore, the MMC can be modeled as a controlled voltage source $\textbf{\textit{u}}_{\text{diff}}$ behind an impedance $R_{\text{s}}+\text{j} \omega L_{\text{s}}$ at the AC terminal, as shown in Fig.~\ref{fig_EquivCircuit}, where the subscripts "on" and "off" denote the onshore and offshore sides, respectively.

For the tractability of converter-grid interaction, Thévenin equivalents are adopted to model the interfaces of the onshore and offshore AC grids, as shown in Fig.~\ref{fig_EquivCircuit}. Specifically, a variable Thévenin voltage source $\textit{\textbf{u}}_{\text{th},k}$ (V) behind a Thévenin equivalent impedance $R_{\text{th},k}+\text{j} \omega L_{\text{th},k}$ ($\Omega$) are modeled, with the parameters referred to the MMC terminal, and $k \in {\{\text{on, off}\}}$.

\begin{figure}[!t]
\centering
\includegraphics{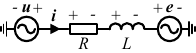}
\caption{Active power transmission through a generic AC network}
\label{fig_URLE}
\end{figure}

The dynamics between the grid interface and MMC at the AC terminal can be modeled using the generic AC network in Fig.~\ref{fig_URLE}, where the AC voltage sources $\textbf{\textit{u}}$ and $\textbf{\textit{e}}$ with amplitudes $U$ (V) and $E$ (V) are connected by a lumped impedance $R+\text{j} \omega L$ ($\Omega$). 
Based on \cite{5308285}, a generic transfer function of the active power dynamics with nonlinear inputs $\alpha_{1}=-L E u^{\text{d}} \cos \! \left(\delta \right)+L E u^{\text{q}} \sin \! \left(\delta \right)+ U^{2} L$; $\alpha_{0}=E \left(L \omega  u^{\text{q}} -R u^{\text{d}} \right) \cos \! \left(\delta \right)+E \left(L \omega  u^{\text{d}} +R u^{\text{q}} \right) \sin \! \left(\delta \right) +U^{2}R$ in dq frame is given by
\begin{align}
\label{NonLinPwr}
    P = \frac{\alpha_{1}s + \alpha_{0} }{(sL+R)^2+( \omega L)^2}, 
\end{align}
where $P$ (W) is the output active power from $\textbf{\textit{u}}$; $s$ is the Laplace operator; $\delta$ (rad) is the power angle (i.e., the angle by which $\textbf{\textit{u}}$ leads $\textbf{\textit{e}}$).

For the onshore and offshore MMCs, respectively, \eqref{NonLinPwr} can be represented as
\begin{subequations}
\begin{align}
    \label{NonLinPwrOn}
    &P_{\text{ac,on}} = P|_{u^{\text{d,q}}=u^{\text{d,q}}_{\text{diff,on}},E=U_{\text{th,on}},R=R_{\text{on}},L=L_{\text{on}},\delta = \delta_{\text{on}}}, \\
    \label{NonLinPwrOff}
        &P_{\text{ac,off}} = P|_{u^{\text{d,q}}=u^{\text{d,q}}_{\text{diff,off}},E=U_{\text{th,off}},R=R_{\text{off}},L=L_{\text{off}},\delta = \delta_{\text{off}}},
\end{align}
\end{subequations}
where $R_{\text{on}} = R_{\text{s}} + R_{\text{th,on}}$; $L_{\text{on}} = L_{\text{s}} + L_{\text{th,on}}$; $R_{\text{off}} = R_{\text{s}} + R_{\text{th,off}}$; $L_{\text{off}} = L_{\text{s}} + L_{\text{th,off}}$; $\delta_{\text{on}} = \theta_{\text{diff,on}} - \theta_{\text{th,on}}$; $\delta_{\text{off}} = \theta_{\text{diff,off}} - \theta_{\text{th,off}}$; $\theta_{\text{diff,on}}, \theta_{\text{diff,off}}, \theta_{\text{th,on}}, \text{and}\, \theta_{\text{th,off}}$ are the phase angles of $\textit{\textbf{u}}_{\text{diff,on}}, \textit{\textbf{u}}_{\text{diff,off}}, \textit{\textbf{u}}_{\text{th,on}}, \text{and}\, \textit{\textbf{u}}_{\text{th,on}}$. 
Therefore, \eqref{NonLinPwrOn} and \eqref{NonLinPwrOff} model the nonlinear dynamics of active power transmission at the AC terminals of onshore and offshore MMCs, where the control inputs are the phase angles and amplitudes (specified as dq components) of $\textbf{\textit{u}}_{\text{diff}}$, as shown in Fig.~\ref{fig_ControlBox}.

\subsubsection{MMC DC terminal}
By subtracting the voltage balancing equations of upper and lower arms, the dynamics of circulating currents $\textbf{\textit{i}}_{\text{c}}$ can be obtained and then simplified to obtain 
\begin{equation}
\label{MMC_CirculatingCurrent}
2L_{\text{a}} \frac {\text{d}i^{j}_{\text{c}}} {\text{d}t} 
+ 2R_{\text{a}} i^{j}_{\text{c}} 
= 
- 2u^{j}_{\text{sum}}
+ U_{\text{dc}},
\quad
\forall j \in {\{\text{a, b, c}\}},
\end{equation}
where $i^{j}_{\text{c}} = ( i^{j}_{l} + i^{j}_{u} ) / 2$; $u^{j}_{\text{sum}}$ is the additive voltage, and $u^{j}_{\text{sum}} = ( u^{j}_{l} + u^{j}_{u} ) / 2$. 
Summing \eqref{MMC_CirculatingCurrent} over phases a, b, and c, the dynamics of $I_{\text{dc}}$ can be obtained and then simplified as 
\begin{equation}
\label{MMC_Idc}
L_{\text{d}} \frac {\text{d}I_{\text{dc}}} {\text{d}t} 
+ R_{\text{d}} I_{\text{dc}} 
= 
- 2u_{\text{sum0}}
+ {U_{\text{dc}}},
\end{equation}
where $I_{\text{dc}} = \sum_{j}^{\text{a,b,c}}i^{j}_{\text{c}}$; $u_{\text{sum0}}$ is the zero-sequence additive voltage, and $u_{\text{sum0}} = \frac{1}{3}\sum_{j}^{\text{a,b,c}}u^{j}_{\text{sum}}$; $L_{\text{d}} = \frac{2}{3}L_{\text{a}}$; 
$R_{\text{d}} = \frac{2}{3}R_{\text{a}}$.

Therefore, the MMC can be modeled as a controlled voltage source with voltage $2u_{\text{sum0}}$ behind an impedance $R_{\text{d}}+\text{j}\omega L_{\text{d}}$ at the DC terminal, as shown in Fig.~\ref{fig_EquivCircuit}. Accordingly, the MMCs' dynamics at the DC port are modeled as a first-order system where $2u_{\text{sum0}}$ is the control input, as shown in Fig.~\ref{fig_ControlBox}.

\subsubsection{MMC Internal energy storage}
Summing the energy stored in the individual submodules, the submodule capacitors can be represented by a single equivalent capacitor as 
\begin{equation}
\label{MMC_energy}
W_{\text{t}}
= 6 N 
\frac{1}{2} C_{\text{SM}}U_{\text{SM}}^2 =\frac{1}{2} C_{\text{eq}}U_{\text{eq}}^2,
\end{equation}
where $W_{\text{t}}$ (J) is the total internal energy of a MMC; $C_{\text{SM}}$ (F) and $U_{\text{SM}}$ (V) are the capacitance and voltage of a single submodule capacitor, and $C_{\text{eq}} = \frac{6}{N}C_{\text{SM}}$ (F) and $U_{\text{eq}}= NU_{\text{SM}}$ (V) are the capacitance and voltage of the equivalent capacitor.

Thus, the total energy dynamics 
\begin{align}
\frac{\text{d}W_{\text{t}}}{\text{d}t} &=  P_{\text{dc}} - P_{\text{ac}} \label{MMC_PowerBalance}
\end{align}
can be equivalently represented by the charge equation
\begin{align}
C_{\text{eq}} \frac {\text{d}U_{\text{eq}}} {\text{d}t} 
&= I_{\text{DCP}} - I_{\text{ACP}} \label{MMC_CeqDynamics}
\end{align}
of the equivalent capacitor. Here $P_{\text{ac}}$ (W) is the AC terminal power, and $P_{\text{ac}} = \sum_{j}^{\text{a,b,c}} u^{j}_{\text{diff}}i^{j}_{\text{s}}$; $P_{\text{dc}}$ (W) is the DC terminal power, and $P_{\text{dc}} = 2u_{\text{sum0}}I_{\text{dc}}$; $I_{\text{DCP}}$ (A) and $I_{\text{ACP}}$ (A) are the equivalent currents flowing through the DC terminal and AC terminal, and $I_{\text{DCP}} = P_{\text{dc}} / U_{\text{eq}}$ and $I_{\text{ACP}} = P_{\text{ac}} / U_{\text{eq}}$.
Therefore, the internal energy storage of an MMC can be modeled as an equivalent capacitor, whose voltage reflects the sum of the submodule voltages that ultimately limit the range of AC and DC voltages $\textbf{\textit{u}}_{\text{diff}}$ and $2u_{\text{sum0}}$ as shown in Fig.~\ref{fig_EquivCircuit}. We adopt \eqref{MMC_PowerBalance} to model the power-energy dynamics in Fig.~\ref{fig_ControlBox}.

\subsection{OWPP}
\begin{figure}[!t]
\centering
\includegraphics[width=\columnwidth]{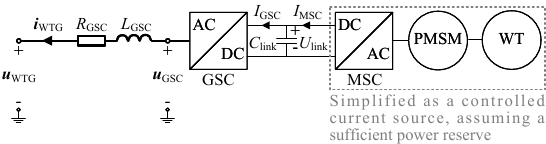}
\caption{Generic type-4 WTG.}
\label{fig_EqWTG}
\end{figure}

A generic model of a type-4 WTG shown in Fig.~\ref{fig_EqWTG} is used to model the OWPP. Here $\textit{\textbf{u}}_{\text{WTG}}$ (V) and $\textit{\textbf{i}}_{\text{WTG}}$ (A) are the phase-to-ground voltage and current at the WTG terminal. The resistance $R_{\text{GSC}}$ ($\Omega$) and inductance $L_{\text{GSC}}$ (H) constitute the GSC filter; $\textit{\textbf{u}}_{\text{GSC}}$ (V) is the phase-to-ground output voltage of the GSC. Finally, $C_{\text{link}}$ (F) is the DC-link capacitor of the equivalent WTG, whose terminal voltage is $U_{\text{link}}$ (V).  

\subsubsection{AC terminal}
According to KVL, the AC voltage balance in Fig.~\ref{fig_EqWTG} is established as 
\begin{equation}
\label{OWPP_GSC}
L_{\text{GSC}} \frac {\text{d}i^{j}_{\text{WTG}}} {\text{d}t} 
+ R_{\text{GSC}} i^{j}_{\text{WTG}} 
= 
u^{j}_{\text{GSC}}
- u^{j}_{\text{WTG}},
\forall j \in {\{\text{a, b, c}\}}.
\end{equation} 
Therefore,  the WTG can be modeled as a controlled voltage source $\textit{\textbf{u}}_{\text{GSC}}$ behind the impedance $R_{\text{GSC}}+\text{j}\omega L_{\text{GSC}}$ at the AC port, as shown in Fig.~\ref{fig_EquivCircuit}, where the subscript index denotes the serial number of the WTG.

Similar to the MMC modeling, the interface
of the OWPP collector network is modeled as a Thévenin equivalent with a variable Thévenin voltage source $\textit{\textbf{u}}_{\text{th,w}}$ (V) behind a Thévenin equivalent impedance $R_{\text{th,w}}+\text{j} \omega L_{\text{th,w}}$ ($\Omega$), as shown in Fig.~\ref{fig_EquivCircuit}. Thus, for the WTG, \eqref{NonLinPwr} can be represented as
\begin{equation}
    \label{NonLinPwrOWPP}
    P_{\text{GSC}} = P|_{u^{\text{d,q}}=u^{\text{d,q}}_{\text{GSC}},E=U_{\text{th,w}},R=R_{\text{w}},L=L_{\text{w}},\delta = \delta_{\text{WTG}}},
\end{equation}
where $P_{\text{GSC}}$ (W) is the output active power of the GSC; $U_{\text{th,w}}$ is the amplitude of $\textit{\textbf{u}}_{\text{th,w}}$; $R_{\text{w}} = R_{\text{GSC}} + R_{\text{th,w}}$; $L_{\text{w}} = L_{\text{GSC}} + L_{\text{th,w}}$; $\delta_{\text{WTG}} = \theta_{\text{GSC}} - \theta_{\text{th,w}}$; $\theta_{\text{GSC}}$ (rad) and $\theta_{\text{th,w}}$ (rad) are the phase angles of $\textit{\textbf{u}}_{\text{GSC}}$ and $\textit{\textbf{u}}_{\text{th,w}}$.

Therefore, \eqref{NonLinPwrOWPP} models the nonlinear dynamics of active power transmission at the AC port of the WTG,
where the control inputs are the phase angle and amplitudes
(specified as dq components) of $\textit{\textbf{u}}_{\text{GSC}}$, as shown in Fig.~\ref{fig_ControlBox}, where the subscript index denotes the serial number of the WTG.

\subsubsection{DC-Link Capacitor}
The DC-Link capacitor $C_{\text{link}}$ is charged and discharged by the MSC and GSC according to
\begin{subequations}
\begin{align}
\frac{\text{d}W_{\text{link}}}{\text{d}t}
= 
P_{\text{MSC}} - P_{\text{GSC}},
\label{OWPP_DCenergy} \\
C_{\text{link}} \frac {\text{d}U_{\text{link}}} {\text{d}t} 
= I_{\text{MSC}} - I_{\text{GSC}},
\label{OWPP_DClink}
\end{align}
\end{subequations}
where $W_{\text{link}}$ (J) is the stored energy of DC-link capacitor, and $W_{\text{link}} = \frac{1}{2}C_{\text{link}}U^2_{\text{link}}$; $P_{\text{MSC}}$ (W) is the output power of MSC; $I_{\text{MSC}}$ (A) and $I_{\text{GSC}}$ (A) are the charging and discharging currents of MSC and GSC, respectively; $I_{\text{MSC}} = P_{\text{MSC}} / U_{\text{link}}$; $I_{\text{GSC}} = P_{\text{GSC}} / U_{\text{link}}$.
Thus, $U_{\text{link}}$ restricts the available $\textit{\textbf{u}}_{\text{GSC}}$ (and the output voltage of the MSC), as shown in Fig.~\ref{fig_EquivCircuit}. Besides, \eqref{OWPP_DCenergy} is adopted to model the power-energy dynamics, as shown in Fig.~\ref{fig_ControlBox}, and \eqref{OWPP_DClink} will be used for RoCoF estimation in the WTG controller (see Fig.~\ref{fig_ControlLaw}(b)).

\subsubsection{Machine-side model}
Assuming a sufficient power reserve and fast torque control through the MSC, the machine-side part (MSC, PMSM, and WT) of the WTG model is simplified as a controlled current source of $I_{\text{MSC}}$, as shown in Fig.~\ref{fig_EquivCircuit}. Equivalently, $P_{\text{MSC}}$ will serve as the control input, as shown in Fig.~\ref{fig_ControlBox}. This simplification is justified given the focus of this work on grid-side control. The simulation studies in Section~\ref{sec:sim} validate the controller using a dynamic WT model.

\subsection{HVDC Line}
The HVDC line has been divided into two $\pi$-sections to accommodate the requirements of the subsequent control design (see Fig.~\ref{fig_EquivCircuit} and Fig.~\ref{fig_ControlBox}), where $R_{\text{dc}}$ ($\Omega$), $L_{\text{dc}}$ (H), and $C_{\text{dc}}$ (F) are the total resistance, inductance, and capacitance  of the HVDC line, respectively; $U_{\text{mid}}$ (V) is the DC voltage at the midpoint of the HVDC line.

\section{Holistic GFM Control for HVDC-OWPPs}

\begin{figure*}[!t]
\centering
\subfloat[]
{\includegraphics[width=\columnwidth]{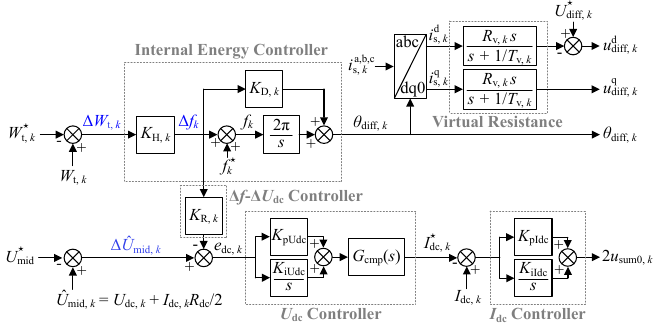}%
\label{fig_MMCcontrol}}
\hfil
\subfloat[]
{\includegraphics[width=\columnwidth]{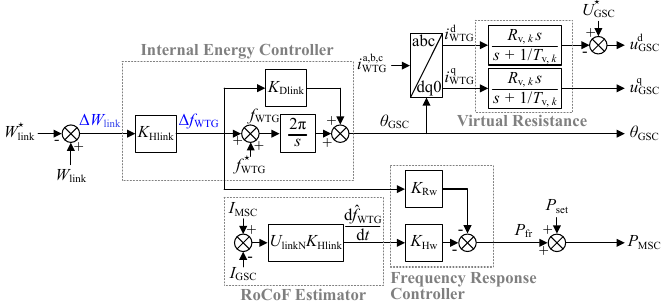}%
\label{fig_OWPPcontrol}}
    \caption{Controllers of the improved holistic GFM control. (a) Controller for the onshore and offshore MMCs ($k\in\{\text{on, off}\}$). (b) Controller for WTGs.}
\label{fig_ControlLaw}
\end{figure*}

The individual controllers used in the proposed holistic GFM control include the onshore and offshore MMCs control as well as the GSC control and MSC power control of the WTGs, as shown in Fig.~\ref{fig_ControlLaw}, where the superscript $\star$ denotes reference values and $\Delta$ denotes deviation. On the time scales of interest, averaged converter models are commonly used and modulation (i.e., NLM and PWM) dynamics are neglected.

\subsection{Onshore and offshore MMC controller}
Both the onshore and offshore MMCs adopt the same controller presented in Fig.~\ref{fig_MMCcontrol}. To begin with, an internal energy controller instead of a power controller is used for power transmission regulation, since the stable internal energy implies the balance of the AC and DC terminal power. A proportional-derivative (PD) controller from energy deviation $\Delta W_{\text{t},k}$ to the frequency deviation $\Delta f_{k}$ (Hz) from the frequency setpoint $f^{\star}_{k}$ (Hz) is used. To avoid computing the time derivative of $\Delta W_{\text{t},k}$, the controller is implemented as proportional-integral (PI) controller from DC-link energy to GSC phase angle. As will be elaborated in the tuning section, \ref{ControlTuning}, the proportional gain $K_{\text{H}, k}$ (Hz/J) links together $\Delta W_{\text{t},k}$ and $\Delta f_{k}$, adjusts the MMC's inertia contribution, and defines the energy regulation bandwidth. The derivative gain $K_{\text{D},k}$ (rad/Hz) results in $\Delta P_{\text{ac},k}$ proportional to $\Delta f_{k}$ (i.e., damping) and defines the phase margin of the closed-loop energy dynamics. Additional damping of circuit dynamics and resonance is provided by adding a transient virtual resistance $R_{\text{v}, k}$ ($\Omega$) with a time constant $T_{\text{v}, k}$ (s)~\cite{5308285}. A standard cascaded AC vector current/voltage control is adopted to track the AC voltage reference. 

To map the frequency variation to the DC voltage, a droop control with gain $K_{\text{R},k}$ (V/Hz) is used. Both onshore and offshore MMCs compute the estimate $\hat{U}_{\text{mid},k}$ of the midpoint voltage of the HVDC line based on their local terminal voltage $\hat{U}_{\text{dc},k}$ and the expected voltage drop resulting from the MMC current injection $I_{\text{dc},k}$. The control objective is to regulate the estimate $\hat{U}_{\text{dc},k}$ to the reference $U^{\star}_{\text{mid}}+K_{\text{R},k} \Delta f_{k}$. PI controllers with proportional and integral gains $K_{\text{pUdc}}$ (A/V), $K_{\text{iUdc}}$ (A/V), $K_{\text{pIdc}}$ (V/A), $K_{\text{iIdc}}$ (V/A) are used for DC voltage and current regulation with zero steady-state error. Moreover, DC line resonances are compensated by $G_{\text{cmp}}(s)$ (see Section~\ref{subsec:dcvolt}).

\subsection{OWPP (WTG) controller}
The controller for WTGs is shown in Fig.~\ref{fig_OWPPcontrol}. Following the same philosophy as the MMC controller design, a PD controller from DC-link energy to frequency is used achieve synchronization and power balancing. The proportional gain $K_{\text{Hlink}}$ (Hz/J) links $\Delta W_{\text{link}}$ and $\Delta f_{\text{WTG}}$, adjusts the inertia of contribution from the DC-link, and defines the energy regulation bandwidth. The derivative gain $K_{\text{Dlink}}$ (rad/Hz) results in $\Delta P_{\text{GSC}}$ proportional to $\Delta f_{\text{WTG}}$ (i.e., damping) and defines the phase margin of the closed-loop energy dynamics. A transient virtual resistance $R_{\text{vw}}$ ($\Omega$) with time constant $T_{\text{vw}}$ (s) is used to suppress the resonance of the dynamics from $\theta_{\text{WTG}}$ to $P_{\text{GSC}}$. To enable  frequency response provision the RoCoF is estimated based on 
\begin{equation}
    \label{rocofWTG}
    \begin{split}
        \frac{\text{d}\Delta f_{\text{WTG}}}{\text{d}t}  &= K_{\text{Hlink}}\frac{\text{d}\Delta W_{\text{link}}}{\text{d}t} = K_{\text{Hlink}}C_{\text{link}}U_{\text{linko}}\frac{\text{d} U_{\text{link}}}{\text{d}t}, \\
        &= K_{\text{Hlink}} U_{\text{linko}} (I_{\text{MSC}} - I_{\text{GSC}}),
    \end{split}
\end{equation}
i.e., without requiring frequency derivatives. Meanwhile, the frequency deviation can be obtained from the DC-link energy controller. Thereby, the frequency response reference $P_{\text{fr}}$ (W) is generated using the gain $K_{\text{Hw}}$ (Ws/Hz) for inertia provision and the gain $K_{\text{Rw}}$ (W/Hz) for FCR provision. Finally, the MSC power command is given sum of the dispatch $P_{\text{set}}$ (W) and the frequency response reference $P_{\text{fr}}$ (W).

\subsection{Steady-state Behavior}
\label{SSbehavior}
According to Fig.~\ref{fig_MMCcontrol}, when the control errors $e_{\text{dc, on}}$ and $e_{\text{dc, off}}$ are regulated to zero, the onshore and offshore frequencies are related by
\begin{equation}
    \label{ssfreq}
    K_{\text{R,on}}\Delta f_{\text{on}} = \Delta \hat{U}_{\text{mid,on}} = \Delta \hat{U}_{\text{mid,off}} = K_{\text{R,off}}\Delta f_{\text{off}}.
\end{equation}
Based on \eqref{ssfreq} and the synchronization at onshore and offshore AC grids, the relationships among the frequencies in the HVDC-OWPP system can be obtained as  
\begin{equation}
    \label{OnOfffreq}
  \Delta f_{\text{WTG}} = \Delta f_{\text{off}} = \frac{K_{\text{R, on}}}{K_{\text{R, off}}} \Delta f_{\text{on}} = \frac{K_{\text{R, on}}}{K_{\text{R, off}}} \Delta f_{\text{th,on}}.
\end{equation}
Therefore, under holistic GFM control, offshore frequency variations  proportionally reflect onshore frequency variations in steady-state, enabling the provision of frequency response service. Crucially, in contrast to the dual-port GFM control in~\cite{10478164}, this relationship holds for a lossy HVDC networks. It is worth noting that very fast convergence to the steady-state, and hence fast provision of inertia and FCR, can be achieved by tuning the $U_{\text{dc}}$ controllers (see Section~\ref{ControlTuning}).

On the other hand, considering Fig.~\ref{fig_ControlBox} and Fig.~\ref{fig_OWPPcontrol}, in steady-state $P_{\text{ac, on}} = P_{\text{dc, on}}$, $P_{\text{ac, off}} = P_{\text{dc, off}}$, and $P_{\text{GSC}} = P_{\text{MSC}}=P_{\text{set}} + P_{\text{fr}}$ must hold.
Neglecting the power loss over the HVDC line and onshore and offshore grids, this results in
\begin{align*}
        \!P_{\text{ac, on}} \!=\! P_{\text{dc, on}} \!=\! P_{\text{dc, off}} \!=\! P_{\text{ac, off}} \!=\! P_{\text{GSC}} \!=\! P_{\text{MSC}} \!=\! P_{\text{set}} \!+\! P_{\text{fr}}
\end{align*}
Thus, the frequency response from OWPP can be transferred to the onshore system without explicit power dispatch at the HVDC converters. Fast convergence to the steady-state can be achieved through tuning of the internal energy controllers and DC-link energy controller.

\subsection{Transient Response}\label{transbehavior}
\begin{figure*}[!t]
\centering
{\includegraphics[width=0.8\textwidth]{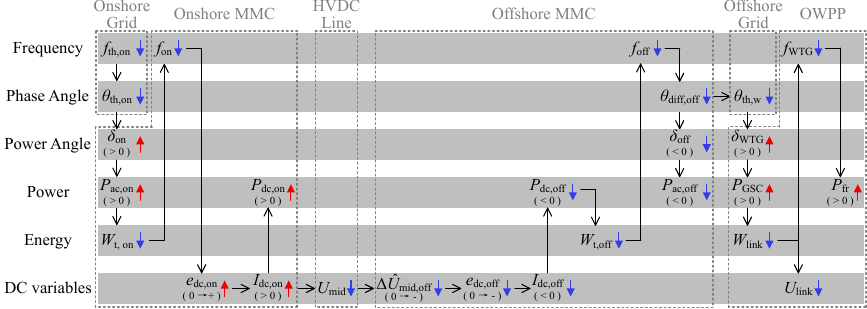}%
\label{fig_TrntBhvFreqDwn}}
\caption{Transient response of the HVDC-OWPP system under the improved holistic GFM control for an under-frequency event.}
\label{fig_TrntBhv}
\end{figure*}

The transient response of the system to an onshore under-frequency event is illustrated in Fig.~\ref{fig_TrntBhv} assuming no deviations before the event.

Therefore, during a transient, the frequencies of the HVDC-OWPP system, $f_{\text{on}}$, $f_{\text{off}}$, and $f_{\text{WTG}}$, deviate in the same direction as $f_{\text{th,on}}$, thereby ensuring appropriate transient frequency response. Meanwhile, the inherent power responses of $P_{\text{ac/dc, on/off}}$ and $P_{\text{GSC}}$ , from the internal energy, $W_{\text{t, on/off}}$ and $W_{\text{link}}$, are supporting the onshore frequency, forming a coherent flow to transfer the frequency response from OWPP.

\section{Controller Tuning}
\label{ControlTuning}
Considering that $K_{\text{Rw}}$ and $K_{\text{Hw}}$ are proportional to the amount of inertia and FCR to be delivered, the tuning of the WT frequency response controller in Fig.~\ref{fig_ControlLaw}b is straightforward. In the following, a systematic tuning method for the virtual resistance, internal energy controller, and DC controllers is introduced.

\subsection{Virtual Resistance}
\label{RvTvTuning}
The robustness and bandwidth of the linearized dynamics $G_{P_{\text{ac}}}(s)$ from power angle to active power with transient virtual resistance have been analyzed in~\cite{8490668}. Neglecting the line resistance ($R$ in Fig.~\ref{fig_URLE}) and assuming large $T_{\text{v}}$, it is suggested that $R_{\text{v}}$ should be $0.2$~p.u. and $T_{\text{v}}$ should be $5/\omega_{\text{N}}$~rad/s to $10/\omega_{\text{N}}$~rad/s, where $\omega_{\text{N}}$ (rad/s) is the nominal angular frequency. Thereafter, $G_{P_{\text{ac}}}(s)$ reduces to a second-order system with a pole pair with natural frequency equal to the system angular frequency $\omega$ (in the remainder, $\omega \approx \omega_{\text{N}}$ is assumed) and damping ratio proportional to $R_{\text{v}}$ and the grid strength ($1/L$ in Fig.~\ref{fig_URLE}), as well as two zeros beyond $\omega_{\text{N}}$ (possibly one in the right-half-plane depending on the operating point).

\subsection{Internal Energy Controller}
\label{nrgBlncTuning}
\begin{figure}[!t]
\centering
\includegraphics{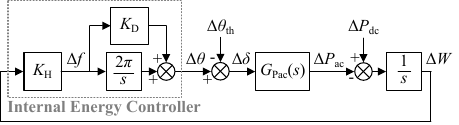}
\caption{Generic linearized closed-loop internal energy regulation dynamics of onshore and offshore MMCs and the GSC of OWPP.}
\label{fig_EngBlncDynamics}
\end{figure}

\begin{figure}[!t]
\centering
\includegraphics{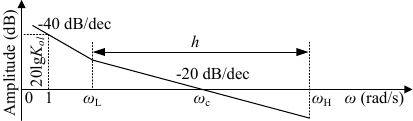}
\caption{Tuning by loop shaping (high-frequency range omitted for brevity).}
\label{fig_TuningBode}
\end{figure}

The linearized dynamics of internal energy regulation of the onshore and offshore MMCs and the WTGs can be generically represented by the system shown in Fig.~\ref{fig_EngBlncDynamics} that resembles the well-known swing equation. 
Therefore, the internal energy $W$ can be regarded as a virtual frequency, and thus, $K_{\text{H}}$ adjusts the virtual inertia and  $K_{\text{D}}$ results in virtual damping, i.e., $\Delta P_{\text{ac}}$ proportional to $\Delta f$. The open loop transfer function $G_{\text{nrg}}(s)$ of the system in Fig.~\ref{fig_EngBlncDynamics} is obtained as 
\begin{equation}
    \label{GolErgBlnc}
G_{\text{nrg}}(s) = K_{\text{H}}\frac{2\pi+K_{\text{D}}s}{s^2}G_{P_{\text{ac}}}(s).
\end{equation}
Using the tuning of $R_{\text{v}}$ and $T_{\text{v}}$ proposed in Section~\ref{RvTvTuning}, only the damped pole pair resonating at $\omega_{\text{N}}$ and the zeros beyond $\omega_{\text{N}}$ remain in $G_{P_{\text{ac}}}(s)$. Using $\omega_{\text{N}}$ as the high-frequency threshold ($\omega_{\text{H}}$), and $2\pi/K_{\text{D}}$ as the low-frequency threshold ($\omega_{\text{L}}$), the Bode diagram of $G_{\text{nrg}}(s)$ is illustrated in Fig.~\ref{fig_TuningBode}. Therefore, the phase margin of the system can be adjusted by tuning $K_{\text{D}}$ for the middle range bandwidth, $h$ ($h=\omega_{\text{H}}/\omega_{\text{L}}$), while the bandwidth of the system can be adjusted by tuning $K_{\text{H}}$ to adjust the open-loop gain and thus, the cross-over frequency  $\omega_{\text{c}}$.

The tuning should focus on the transfer of both frequency variation and power response.
To specify the tuning targets,
\begin{subequations}
\begin{align}
    \label{FreqTracking}
     &y_{\text{ac-ac}}(t) = \mathcal{L}^{-1} \left\{ \frac{1}{s^2} \frac{\Delta W(s)}{\Delta \theta_{\text{th}}(s)} \right\}, \\
    \label{PwrTransfer}
     &y_{\text{dc-ac}}(t) = \mathcal{L}^{-1} \left\{ \frac{1}{s} \frac{\Delta P_{\text{ac}}(s)}{\Delta P_{\text{dc}}(s)} \right\},
\end{align}
\end{subequations}
are defined based on the inverse Laplace transform $\mathcal{L}^{-1}$ of the ramp response of $\Delta W$ to $\delta \theta_{\text{th}}$ and the step response of $\Delta P_{\text{ac}}$ to $\Delta P_{\text{dc}}$. The ramp response $y_{\text{ac-ac}}$ reflects the dynamics of both the frequency variation transfer and the active power interactions at the onshore and offshore AC grids, since $W$ can serve as a virtual frequency and indicate the power balance at the same time. The step response $y_{\text{dc-ac}}$ models the power transfer between the DC and AC terminals of the converter.
Setting $\omega_{\text{c}}$ in the middle of $\omega_{\text{L}}$ and $\omega_{\text{H}}$ on the logarithmic axis balances the performance indices (i.e., the overshoot, peak time, and settling time of $y_{\text{ac-ac}}$ and $y_{\text{dc-ac}}$). Introducing the bandwidth $h$ (see Fig.~\ref{fig_TuningBode}) this choice results in the gains
\begin{equation}
    \label{KHsetting}
K_{\text{H}} = \frac{\omega^2_{\text{N}}}{2 \pi G_{P_{\text{ac}}}(0) \sqrt{h^{3}}}, \quad K_{\text{D}} = \frac{2\pi h}{\omega_{\text{N}}},
\end{equation}
and tuning $K_{\text{H}}$ and $K_{\text{D}}$ is reduced to tuning $h$. A small $h$ ensures faster energy regulation, while a large $h$, e.g. larger than $5$, ensures a phase margin of about $30$~deg.

Overall, the transfer bandwidth of the frequency variation and power response at an AC terminal and between the AC and DC terminals of a converter is bounded by the system angular frequency $\omega_{\text{N}}$. As $h$ is suggested to be larger than $5$ to achieve at least $30$~deg phase margin, the largest bandwidth is expected to be around $\omega_{\text{N}}/\sqrt{5}$.
\subsection{DC Current and Voltage Controller}

\subsubsection{DC Current Controller}
To obtain a stable pole-zero cancellation, the gains of the DC current controller of both onshore and offshore MMCs are selected as
\begin{equation}
    \label{DCcurrentSetting}
    K_{\text{pIdc}} = L_{\text{d}} \omega_{\text{idc}}, \quad K_{\text{iIdc}} = R_{\text{d}} \omega_{\text{idc}}.
\end{equation}
In this way, the inner current loop is simplified as a first-order system with bandwidth $\omega_{\text{idc}}$ (rad/s), which is limited by sampling frequency $\omega_{\text{s}}$ (rad/s). A useful rule of thumb is $\omega_{\text{idc}} \leq \omega_{\text{s}}/10$~\cite{sharifabadi2016design}.

\subsubsection{DC Voltage Controller}\label{subsec:dcvolt}
\begin{figure}[!t]
\centering
\includegraphics[width=\columnwidth]{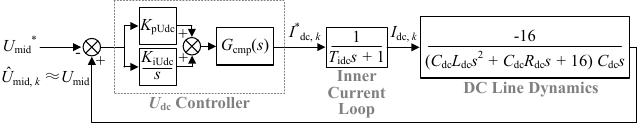}
\caption{DC voltage control of the onshore or offshore MMC ($k\in\{\text{on, off}\}$).}
\label{fig_OuterVoltageLoop}
\end{figure}

After tuning the inner current loop, the reduced model in Fig.~\ref{fig_OuterVoltageLoop} can be used for voltage loop tuning. Here $T_{\text{idc}} = 1/\omega_{\text{idc}}$ is the time constant of the closed-loop DC current dynamics. The DC line dynamics introduce a pole pair resonating at $\sqrt{16/C_{\text{dc}}L_{\text{dc}}}$. To damp this resonance, a compensator 
\begin{align}
    \label{compensation}
    G_{\text{cmp}(s)} = \frac{C_{\text{dc}}L_{\text{dc}}s^2+C_{\text{dc}}R_{\text{dc}}s+16}{16(0.1s/\omega_{\text{idc}}+1)^2}
\end{align}
can be designed so that the resonating pole pair is replaced by one with faster dynamics and larger damping ratio. After compensation, the open-loop transfer function of the DC voltage control system can be simplified to
\begin{align}
    \label{DCvoltageOpenLoop}
    G_{\text{Udc}}(s) = \frac{K_{\text{iUdc}} (\frac{K_{\text{pUdc}}}{K_{\text{iUdc}}}s+1)}{C_{\text{dc}} s^2 (s/\omega_{\text{idc}}+1)(0.1s/\omega_{\text{idc}}+1)^2}.
\end{align}
The tuning objective is to ensure stability as well as fast reference tracking and disturbance rejection. As discussed in Section~\ref{SSbehavior} and Section~\ref{transbehavior}, the onshore frequency variation is first introduced into $e_{\text{dc,on}}$, thereby shifting the onshore DC voltage reference. Then, the action of the $U_{\text{dc,on}}$ controller results in a deviation of $U_{\text{dc,off}}$ and $W_{\text{t,off}}$ from their setpoints. The $U_{\text{dc,off}}$ controller and deviation of $W_{\text{t,off}}$ maps these deviations to a change in $e_{\text{dc,off}}$ (i.e., the offshore DC reference). Finally, the power response from offshore side is perceived as a disturbance to the $U_{\text{dc,on}}$ controller, and the transfer of the power response is achieved through the disturbance rejection by the $U_{\text{dc,on}}$ controller. Therefore, the transfer of frequency variation and power response is coupled through the combined mechanism of reference tracking and disturbance rejection in the DC control system.

Using $\omega_{\text{L}} = K_{\text{iUdc}}/K_{\text{pUdc}}$ and $\omega_{\text{H}} = \omega_{\text{idc}}$, the tuning proposed in Section~\ref{nrgBlncTuning} and illustrated in Fig.~\ref{fig_TuningBode} can be adopted for the DC voltage controller tuning. The gains
\begin{equation}
    \label{UdcPIsetting}
    K_{\text{pUdc}} = \frac{C_{\text{dc}}\omega_{\text{idc}}}{\sqrt{h}}, \quad K_{\text{iUdc}} = \frac{C_{\text{dc}}\omega_{\text{idc}}^2}{\sqrt{h^{3}}},
\end{equation}
result in the bandwidth $h$ and $\omega_{\text{c}}$ in the middle between $\omega_{\text{L}}$ and $\omega_{\text{H}}$ on the logarithmic axis to balance the performance in reference tracking and disturbance rejection. Using this approach, tuning $K_{\text{pUdc}}$ and $K_{\text{iUdc}}$ is reduced to tuning $h$ only. A small $h$ ensures faster DC regulation, while a large $h$, e.g. larger than $4$, ensures a phase margin of about $30$~deg.

Thus, the bandwidth of frequency transfer and power response in the DC control system is bounded by $T_{\text{idc}}$ and ultimately $\omega_{\text{s}}$. As $h$ is suggested to be larger than $4$ to achieve at least $30$~deg phase margin, the largest bandwidth is expected to be around $\omega_{\text{s}}/20$.

\subsection{Frequency - DC voltage droop}
$K_{\text{R},k}$ can be set in terms of the allowable variation range of frequency and DC voltage as
\begin{equation}
    \label{fDCdroop}
    K_{\text{R},k} = \frac{\Delta U_{\text{dcm}}}{\Delta f_{\text{m}, k}}, \quad k \in \{ \text{on, off} \},
\end{equation}
where $\Delta U_{\text{dcm}}$ (V) is the maximum allowable deviation of the midpoint DC voltage and $\Delta f_{\text{m}, k}$ (Hz) is the maximum allowable deviation of onshore/offshore grid frequency.


\section{Simulation Studies}\label{sec:sim}
In this section, the inertial response and FCR delivery of the proposed holistic GFM control is illustrated through simulation studies. Furthermore, its performance is compared with representative state-of-the-art control configurations.

\subsection{Benchmark System}
Electromagnetic transient simulations of a load step in the system shown in Fig.~\ref{fig_SysScheme} are conducted using MATLAB/Simulink and an Euler integration method with $50\mu\text{s}$ time step. The onshore system is modeled by an equivalent synchronous generator. The OWPP is aggregated into one single equivalent WTG \cite{1708945} to increase simulation efficiency, and the machine-side dynamics are modeled based on a model of the GE $1.5$~MW WTG~\cite{miller2003dynamic}. The specific models and the corresponding parameters are summarized in Table~\ref{tab_model_params}

\begin{table}[h!]
\centering
\caption{Models and parameters for the benchmark system}
\begin{tabular}{|m{0.3\columnwidth}| p{0.6\columnwidth}|}
\hline
\textbf{} & \multicolumn{1}{c|}{\textbf{Model \& Parameters}}\\ 
\hline
\textbf{Onshore System} &
Onshore equivalent synchronous machine, transformer, and AC line are taken from the two-area system~\cite[p.~813]{kundur2007power} with the inertia modified to $2$~s and $5\%$ frequency droop. \\[4pt]
\hline
\textbf{Onshore \& Offshore MMC and HVDC line} &
Arm-average MMC model, transformer, and frequency-dependent HVDC line from \cite{9729620}. \\[4pt]
\hline
\textbf{Offshore AC grid} &
$\pi$-section model with $R = 0.028~\Omega/\text{km}$, $L = 0.45~\text{mH}/\text{km}$, $C = 0.32~\mu\text{F}/\text{km}$. \\[4pt]
\hline
\textbf{OWPP} &
Single equivalent WTG with $R_{\text{GSC}} = 0.019~\Omega$, $L_{\text{GSC}} = 3~\text{mH}$, 
$C_{\text{link}} = 5.17~\text{mF}$, $U_{\text{link}} = 132~\text{kV}$, and the ”Wind Turbine \& Turbine
Control Model” in~\cite{miller2003dynamic} with constant wind speed of $12$~m/s and rotor speed regulated at the nominal speed.\\ \hline
\end{tabular}
\label{tab_model_params}
\end{table}

\subsection{Control Strategies}
The holistic GFM is compared with the main approaches available in the literature, as listed in Table~\ref{table_CtrlConfig}, where $H_{\text{cvtr}}$ represents the equivalent inertia stored in the inherent energy storage of the converter. 

\subsubsection{Classic GFL} the HVDC converter controls from~\cite{CIGRE_604_2014} for HVDC and WTG controls from~\cite{noauthor_model_2023} are  implemented and their parameters are adapted to the benchmark system. 
\subsubsection{GFM OWPP}
Starting from the classic GFL configuration, the GSC control is replaced by dual-port GFM control using DC-voltage synchronization~\cite{10734179}. The controller is tuned to deliver an inertia response according to the energy in $C_{\text{link}}$ under the optimal damping ratio of $0.707$. 
\subsubsection{GFM HVDC}
Starting from the classic GFL configuration, the onshore MMC control is replaced by dual-port GFM control using internal-energy-based synchronization \cite{9887878}. The controller is tuned to deliver an inertia response according to the energy $W_{\text{on}}$ under the optimal damping ratio of $0.707$. 
\subsubsection{Full AC-GFM}
This configuration combines the GFM OWPP and GFM HVDC configurations. 
\subsubsection{Improved Holistic GFM} The proposed control is tuned to maximize the control bandwidth. Consequently, the inertia delivered from the internal converter energy storage is minimized. This tuning trades off the relatively small available inertia from the inherent energy storage of converters for fast transfer of frequency variation and power response. The control parameters are summarized in Table~\ref{table_HolisticParams}.

\begin{table*}[!t]
\caption{Control configurations for performance comparison}
\centering
\begin{tabular}{|c|c|c|c|c|}
\hline
 & \textbf{Approach} & \textbf{Onshore MMC} & \textbf{Offshore MMC} & \textbf{OWPP (WTG)}\\
\hline
\textbf{Classic GFL} & \makecell{CBC\\Delay: 50 ms} & GFL & $V-f$ GFM & GFL\\
\hline
\textbf{GFM OWPP} & CFC & GFL & $V-f$ GFM & \makecell{Dual-port GFM\\$H_{\text{cvtr}}$ provided: 100\%} \\
\hline
\textbf{GFM HVDC} & CFC & \makecell{Dual-port GFM\\$H_{\text{cvtr}}$ provided: 100\%} & $V-f$ GFM & GFL\\
\hline
\textbf{Full AC-GFM} & CFC & \makecell{Dual-port GFM\\$H_{\text{cvtr}}$ provided: 100\%} & $V-f$ GFM & \makecell{Dual-port GFM\\$H_{\text{cvtr}}$ provided: 100\%}\\
\hline
\textbf{Improved Holistic GFM} & CFC & \makecell{Dual-port GFM\\$H_{\text{cvtr}}$ provided: 65\%} & \makecell{Dual-port GFM\\$H_{\text{cvtr}}$ provided: 54\%} & \makecell{Dual-port GFM\\$H_{\text{cvtr}}$ provided: 7\%} \\
\hline
\end{tabular}
\label{table_CtrlConfig}
\end{table*}

\begin{table}[!t]
\caption{Parameters of the improved holistic GFM control}
\centering
\begin{tabular}{|c|c|c|}
\hline
\multicolumn{3}{|c|}{\textbf{Onshore \& Offshore MMCs} ($h_{\text{ac}}=5, h_{\text{dc}}=4$)} \\
\hline
$\!\!K_{\text{H,on}}$ (Hz/J): $1.57\!\!\times\!\!10^{-6}$\!\! & $\!\!K_{\text{D,on}}$ (rad/Hz): $0.1$\!\! & $\!\!K_{\text{R,on}}$ (V/Hz): $3.84\!\!\times\!\!10^4$\!\! \\
\hline
$\!\!K_{\text{H,off}}$ (Hz/J): $1.89\!\!\times\!\!10^{-6}$\!\! & $\!\!K_{\text{D,off}}$ (rad/Hz): $0.1$\!\! & $\!\!K_{\text{R,off}}$ (V/Hz): $3.84\!\!\times\!\!10^4$\!\! \\
\hline
\multicolumn{3}{|c|}{
\begin{tabular}{ c|c|c|c }
$R_{\text{v,on}} (\Omega): 46$ & $T_{\text{v,on}}$ (s): $0.021$ & $R_{\text{v,off}} (\Omega): 52$ & $T_{\text{v,off}}$ (s): $0.017$
\end{tabular}
} \\
\hline
$K_{\text{pUdc}}$ (A/V): $0.024$ & $K_{\text{iUdc}}$ (A/V): $6.1$ & $K_{\text{pIdc}}$ (V/A): $130$ \\
\hline
\multicolumn{3}{|c|}{
\begin{tabular}{ c|c }
 $K_{\text{iIdc}}$ (V/A): $2048$ & $G_{\text{cmp}(s)} = \frac{3.5\!\!\times\!\!10^{-6}s^2+2.1\!\!\times\!\!10^{-4}s+16}{16(1\!\!\times\!\!10^{-4}s+1)^2}$
\end{tabular}
} \\
\hline
\multicolumn{3}{|c|}{\textbf{OWPP} ($h_{\text{ac}}=15$)} \\
\hline
\multicolumn{3}{|c|}{
\begin{tabular}{ c|c }
$K_{\text{Hlink}}$ (Hz/J): $1.1\!\!\times\!\!10^{-6}$ & $K_{\text{Dlink}}$ (rad/Hz): $0.31$
\end{tabular}} \\
\hline
\multicolumn{3}{|c|}{
\begin{tabular}{ c|c }
$R_{\text{vw}} (\Omega): 2.4$ & $T_{\text{vw}}$ (s): $0.015$
\end{tabular}
} \\
\hline
\end{tabular}
\label{table_HolisticParams}
\end{table}

\subsection{Simulation Results and Discussion}
\subsubsection{FCR and Inertial Response Provision by the proposed Holistic GFM Control}
\begin{figure}[!t]
\centering
\subfloat[]
{\includegraphics[width=\columnwidth]{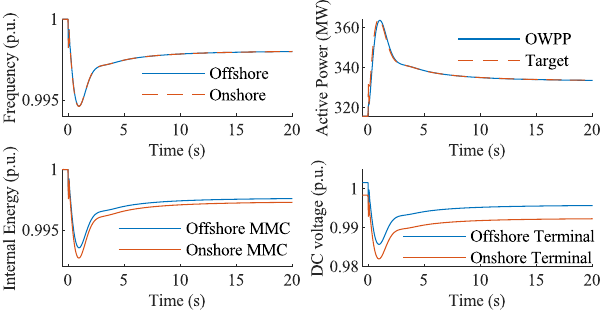}%
\label{fig_FCRprovision}}
\hfil
\subfloat[]
{\includegraphics[width=\columnwidth]{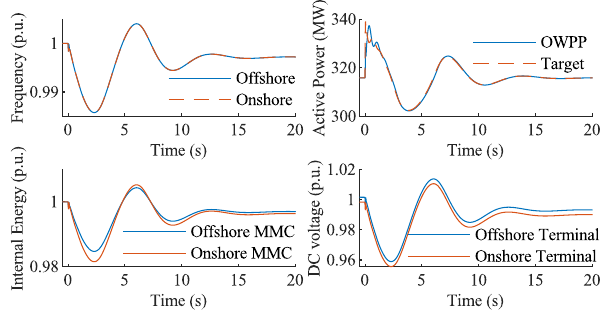}%
\label{fig_IRprovision}}
\caption{Simulation results of FCR and inertial response provision by the HVDC-OWPP system using the proposed holistic GFM control. (a) FCR provision ($5\%$ frequency droop); (b) Inertial response provision ($2H_{\text{OWPP}}=4$~s).}
\label{fig_FCRnIRprovision}
\end{figure}

The FCR and inertial response provision by the HVDC-OWPP system using the proposed holistic GFM control with the proposed tuning are shown in Fig.~\ref{fig_FCRnIRprovision}. The target 
active power response (dashed) of the OWPP shown in the active power plot in Fig.~\ref{fig_FCRnIRprovision} is generated by multiplying the onshore frequency deviation and RoCoF by the target FCR droop constant and inertia constant, respectively. In both cases, the onshore frequency variation is transferred to the offshore side fast and accurately, as the onshore and offshore frequencies almost overlap; the OWPP's active power follows the reference closely; the internal energies follow the frequency deviation proportionally with minor differences between each other due to using different gains $K_{\text{H},k}$ resulting from different Thévenin impedance to which the MMCs are connected; DC voltage deviations also follow the frequency deviation almost proportionally. The overall response matches both the steady-state and transient behaviors described in Section~\ref{SSbehavior} and Section~\ref{transbehavior}.

\subsubsection{Comparison with State-of-the-Art Control configurations}
\begin{figure}[!t]
\centering
\subfloat[]
{\includegraphics[width=\columnwidth]{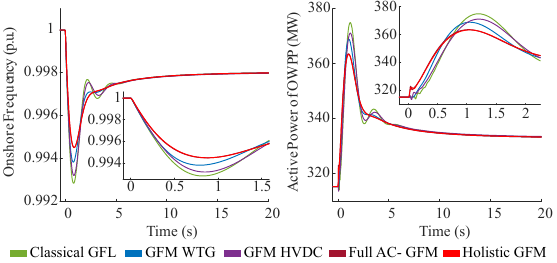}%
\label{fig_FCRcomparison}}
\hfil
\subfloat[]
{\includegraphics[width=\columnwidth]{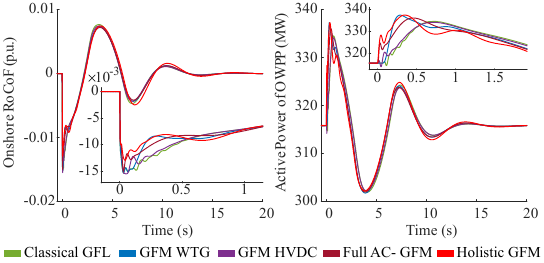}%
\label{fig_IRcomparison}}
\caption{Performance comparison among the proposed holistic GFM control and the representative state-of-the-art control configurations. (a) FCR provision ($5\%$ frequency); (b) Inertial response provision ($2H_{\text{OWPP}}=4$~s).}
\label{fig_FCRnIRcomparison}
\end{figure}

Simulation results for all aforementioned controls are shown in Fig.~\ref{fig_FCRnIRcomparison}. Compared to other control configurations, the improved holistic GFM control reduces the maximum onshore frequency deviation by more than $20\%$ for FCR provision and the maximum onshore RoCoF by more than $15\%$ for the inertial response case.

When providing FCR, the holistic GFM control achieves only slightly less onshore frequency deviation than the full AC-GFM control does. However, this is achieved by reducing the inertia contribution from the MMCs' internal energy and WTG DC link by $40\%$ ,i.e., enabling more reliable operation of the converters and/or reducing hardware cost.

The faster response speed using the holistic GFM control makes more significant differences in the case of inertia provision. In this case, the maximum onshore RoCoF is improved by more than $13\%$ compared to the full AC-GFM configuration which achieves the second best performance.

Therefore, holistic GFM control tuned to maximize the bandwidth (i.e., minimize the converter inertia contribution) prioritizes the fast transfer capability of converters over the utilization of their inherent energy storage. This enables more effective and reliable converter operation with faster frequency response and consequently improved frequency support.

\section{Conclusion}
In this paper, an improved holistic GFM control for HVDC-OWPP systems has been proposed to enhance the provision of FCR and inertial response, with a corresponding analytical control tuning method. In the proposed control architecture, GFM controls are implemented at all AC and DC terminals of an HVDC-OWPP system and coordinated without communication. The proposed tuning method enables a simplified tuning procedure on the middle range bandwidth alone and identified feasible bandwidth boundaries. However, the overall response characteristics of the entire HVDC-OWPP system remain to be studied. 

Simulation results verified that the holistic GFM control enables the HVDC-OWPP system to provide both accurate FCR and inertial response that closely matches ideal targets. Compared to other control configurations, the proposed holistic GFM control can improve the maximum onshore frequency deviation by more than $20\%$ for FCR provision and the maximum onshore RoCoF by more than $15\%$ for the inertial response for the scenario under consideration. At the same time, the utilization of converters' inherent energy storage is minimized. The comparison supports a new converter control design philosophy that prioritizes fast transfer capability over using the converters' inherent energy storage for inertia emulation.





\bibliography{IEEEabrv,Refs}
\bibliographystyle{IEEEtran}

 




\vfill

\end{document}